\documentclass[10pt,conference]{IEEEtran}
\IEEEoverridecommandlockouts
\usepackage{cite}
\usepackage{amsmath,amssymb,amsfonts}
\usepackage{algorithmic}
\usepackage{graphicx}
\usepackage{textcomp}
\usepackage{xcolor}
\usepackage{makecell}
\usepackage{multirow}
\usepackage{amsmath}
\usepackage{threeparttable}
\def\BibTeX{{\rm B\kern-.05em{\sc i\kern-.025em b}\kern-.08em
    T\kern-.1667em\lower.7ex\hbox{E}\kern-.125emX}}

\begin{document}


\title{Data Balancing Improves Self-Admitted Technical Debt Detection}


\author{\IEEEauthorblockN{Murali Sridharan\textsuperscript{1}, Mika Mantyla\textsuperscript{2}, Leevi Rantala\textsuperscript{3}, Maelick Claes\textsuperscript{4}}
\IEEEauthorblockA{\textit{M3S, ITEE}\\
\textit{University of Oulu}\\
Oulu, Finland\\
\{murali.sridharan\textsuperscript{1},mika.mantyla\textsuperscript{2},leevi.rantala\textsuperscript{3},maelick.claes\textsuperscript{4}\}@oulu.fi}
}

\maketitle

\begin{abstract}

A high imbalance exists between technical debt and non-technical debt source code comments. Such imbalance affects Self Admitted Technical Debt (SATD) detection performance, and existing literature lacks empirical evidence on the choice of balancing technique.
In this work, 
we evaluate the impact of multiple balancing techniques, including Data level, Classifier level, and Hybrid, for SATD detection in Within-Project and Cross-Project setup. 
Our results show that the Data level balancing technique SMOTE or Classifier level Ensemble approaches with Random Forest or XGBoost are reasonable choices depending on whether the goal is to maximize Precision, Recall, F1, or AUC-ROC.
We compared our best-performing model with the previous SATD detection benchmark (cost-sensitive Convolution Neural Network). Interestingly the top-performing XGBoost with SMOTE sampling improved the Within-project F1 score by 10\% but fell short in Cross-Project set up by 9\%. This supports the higher generalization capability of deep learning in Cross-project SATD detection, yet while working within individual projects, classical machine learning algorithms can deliver better performance. We also evaluate and quantify the impact of duplicate source code comments in SATD detection performance. Finally, we employ SHAP and discuss the interpreted SATD features.
We have included the replication package\footnote{ https://figshare.com/s/87a4b5002c7488822e60} and shared a web-based SATD prediction tool\footnote{ https://balancing-technical-debt.herokuapp.com/} with the balancing techniques in this study. 

\end{abstract}


\section{Introduction}
Software development is often hampered by time-pressure. The quick-fix mentality that focuses merely on the immediate goal and short-term benefit has been the norm~\cite{kuutila2020time}. 
The quick fixes often turn out to be sub-optimal, as they lack a holistic approach to software maintenance making the source code more rigid without room for future enhancements. They incur a substantial cost in terms of time and man-effort to refactor the code at a later stage. Such accumulated debt by choosing quick delivery over quality delivery is known as \emph{technical debt} in Software Engineering.

The early detection of such technical debt would be instrumental in reducing the increased software maintenance cost. The established approach among the software practitioners have been to utilize static code analysis for improving code quality. Often, the developer acknowledged or developer induced hacky patch/workaround goes unnoticed. Such patch/workaround are expressed by the software developers through source code comments. These are termed as Self-Admitted Technical Debt comments by Potdar et al.~\cite{potdar2014exploratory}. Such SATD comments from the source code has vital information about the source code segments that needs refactoring. 

The detection of technical debt from SATD source code comments has gathered significant interest in the recent past. A crucial challenge associated with technical debt detection from source code comments is the imbalanced distribution among SATD and non-SATD data instances. The supervised machine learning approaches expect a uniform distribution among the data samples for optimal prediction performance but quite often end up with imbalanced data which will affect the prediction capability. Longadge et al.~\cite{longadge2013class} state that imbalanced/skewed data increase the False Negatives (FN), which will decrease the Recall of the minor class (SATD comments in our context). Thabtah et al.~\cite{thabtah2020data} highlight the varying effect of data imbalance for each evaluation metric particularly Precision, Recall and ROC-AUC. The challenge of generating more accurate results while accounting for the class imbalance in the training data is paramount for reliable inference. In reality, this data imbalance problem is very common among multiple tasks in software engineering domain and other domains as well. Previous work on technical debt detection discusses different machine learning approaches but very few have employed balancing techniques for addressing the class imbalance. For 
example Pecorelli et al.~\cite{pecorelli2019comparing} used data sampling on code metrics for detecting code smells and Ren et al.~\cite{ren2019neural} used classifier level balancing technique COST to improve SATD detection from source code comments. 


To the best of our knowledge, we have performed the first extensive empirical study on evaluating multiple balancing techniques for technical debt detection from source code comments. These include data-level balancing techniques (SMOTE, ADASYN, BorderLine SMOTE and SVMSMOTE), classifier-level balancing techniques (COST and Ensemble) and hybrid balancing techniques that combine either data-level and classifier-level balancing techniques or employ customised combined algorithms, for example COST based ensemble algorithm or sampling based ensemble, for handling data imbalance scenarios.

We empirically evaluate three categories of balancing techniques and their impact on technical debt detection from source code comments using regression, bagging and boosting classifiers. More specifically, we compare the performance, in terms of precision, recall, F1 score and area under the ROC curve (ROC-AUC), against the BASELINE approach which employs the same classifier without COST or data sampling balancing techniques. Our results enable software engineering researchers and practitioners to choose relevant balancing technique depending on the use case and the evaluation metric in focus.

In this paper, we answer the following research questions:
\begin{itemize}
    \item \textbf{Balancing the Imbalanced Data: Main RQ} Which balancing technique contribute to better Technical Debt Detection in highly imbalanced source code comments data?
    \begin{itemize}
    \item \textbf{RQ 1 Precise detection of SATD comments:} Which balancing technique consistently contribute to better precision in classifying SATD comments?
    \item \textbf{RQ 2 Extensive detection of SATD comments:} Which balancing technique consistently contribute to better recall while classifying SATD comments?
    \item \textbf{RQ 3 Improving distinction capability of classifier for SATD detection:} Which balancing technique consistently improve the distinction capability of the machine learning model (ROC-AUC) to distinguish between SATD and non-SATD comments?    
    \item \textbf{RQ 4 Balancing Precision and Recall for SATD Detection:} Which balancing technique consistently improve the overall classification performance (F1) for detecting SATD comments?
    \end{itemize}
\end{itemize}

Our main contributions in this paper include:
\begin{itemize}
    \item Evaluation of multiple balancing schemes for SATD detection from highly imbalanced source code comments data.
    \item Recommendations for choosing a balancing technique for SATD detection through source code comments, depending on the evaluation metric in focus.    
    \item Impact analysis of data level, classifier level and hybrid  balancing techniques on evaluation metrics such as Precision, Recall, F1 and ROC-AUC scores.
    \item Web-based SATD detection tool in batch and online modes based on the data from 10 open source projects.
    \item Replication package of our experiments for verification and further extension.    
\end{itemize}

The rest of this paper is organized as follows. First, in Section~\ref{sec:background}, we discuss the evolution of TD and the past works associated with the detection of technical debt. In Section~\ref{sec:methodology}, we discuss the classifiers, data characteristics, and the different balancing techniques studied in this work. We tabulated and discussed our experimental results in Section~\ref{sec:results}.
In Section~\ref{sec:quality}, we discuss the technique used for feature interpretation and the impact of balancing techniques on features. We discuss the implications associated with choosing an appropriate balancing technique in Section~\ref{sec:discussion}. The potential factors that could mine the validity of our results are elaborated in Section~\ref{sec:threats}. We conclude with our main findings from this study in Section~\ref{sec:conclusion} followed by acknowledgements.

\section{Background}\label{sec:background}
\subsection{Technical Debt}
The Technical Debt (TD) metaphor originates from the early 90's, when Cunningham said that shipping immature code resembles like taking a debt~\cite{cunningham1992wycash}. 
TD as a category does not represent a uniform type of debt. Previous study by Alves et al.~\cite{alves2014towards} lists 13 categories for TD including architectural debt, code debt, design debt and people debt. The most recent definition for TD can be found from Avgeriou et al.'s.~\cite{avgeriou2016managing} work, where it is defined as:

\begin{quote}
In software-intensive systems, technical debt is a collection of design or implementation constructs that are expedient in the short term, but set up a technical context that can make future changes more costly or impossible. Technical debt presents an actual or contingent liability whose impact is limited to internal system qualities, primarily maintainability and evolvability.
\end{quote}

In 2014, Potdar et al.~\cite{potdar2014exploratory} coined the term self-admitted technical debt (SATD). SATD is a subset of TD, where the developers have left a note admitting that they intentionally incurred TD. The software practitioners leave comments in the source code which is crucial for detecting SATD in the source code. These have two significant benefits, firstly, it is easier and more efficient to detect SATD from source code, and it also does not need to rely on predefined metrics which are difficult to determine as suggested by Maldonado et al.~\cite{da2017using}.


\subsection{Machine Learning in SATD Detection}

Zampetti et al.~\cite{zampetti2017recommending} studied how five machine learning techniques were able to recommend the software practitioners when to admit design debt using structural and readability metrics from source code and warnings from static analysis tools. They achieved an average precision 0.5 and recall 0.52 for SATD detection within project. For cross-project SATD detection they achieved 0.67 precision and 0.55 recall. Unlike them, we do not study different code metrics, but rather evaluate the impact of multiple balancing schemes using different machine learning techniques when detecting SATD from source code comments.

Maldonado et al.~\cite{da2017using}, developed a maximum entropy classifier for detecting design and requirement SATD from code comments. The classifier builds a maximum entropy model, which is described to be equivalent to multi-class regression model. Their data set consist of 10 projects, which represent different application domains. They used leave-one-out cross-project validation, which means that the classifier was trained on 9 projects and tested against the one which was left out. They achieved an average F1 of 0.620 (min. 0.470, max. 0.814) for design SATD, and F1 of 0.403 (min. 0.154, max. 0.804) for requirement SATD. 

In another related work by Huang et al.~\cite{huang2018identifying}, they developed a machine learning model, which relied on several sub-classifiers. Each of the sub-classifiers used Naive Bayes (NB) multinomial with feature selection technique, and the classification of whether a code comment has SATD or not was based on majority voting done by these classifiers. The data set consisted of 8 different open source projects. Each sub-classifier was trained with leave-one-out cross-project validation, where they used 7 projects for training, and  left one project out for testing. Each sub-classifier left out a different project. They achieved an average F1 of 0.737 (min. 0.518, max. 0.841). 

Ren et al.~\cite{ren2019neural}, trained a Convolution Neural Network (CNN) for SATD detection from source code comments using data from of 10 open source Java projects. They evaluated their approach in Within and Cross-project setup. For Within-project they obtained an average F1 of 0.752 (min. 0.445, max. 0.932), and for Cross-Project the average F1 was 0.766 (min. 0.599, max. 0.878).

\subsection{Imbalanced Data} 
Technical Debt detection through source code comments suffers from imbalanced data problem. Besides SATD detection~\cite{ren2019neural}, it has also been investigated on code smell detection~\cite{pecorelli2019comparing, pecorelli2019role}.

Ren et al.~\cite{ren2019neural}, in their work experiment with normal and weighted cross-entropy loss functions. For addressing imbalance in data, they employed COST (assigning class weight). For weight calculation, $n$ represent the total number of SATD comments, and $m$ the total amount of non-SATD comments
\begin{itemize}
    \item SATD Comment Weight $= \frac{n}{n+m}$
    \item Non-SATD Comment Weight $= \frac{m}{n+m}$
\end{itemize}
The CNN with weighted cross-entropy loss function produced better results than the CNN using normal cross-entropy loss function. The authors note that the effectiveness of the weighted cross-entropy loss function increased based on the imbalance in the data~\cite{ren2019neural}.

Huang et al.~\cite{huang2018identifying} employed feature selection technique to extract features from source code comments during training. They extracted features from 8 open source projects and combined multiple classifiers into a composite classifier for SATD comments detection task. They claimed a performance improvement of 409\% over the work by Potdar et al.\cite{potdar2014exploratory}.

In~\cite{pecorelli2019role}, the authors investigated five different data balancing methods when detecting code smells with source code metrics. They examined 125 releases from 13 open source systems from a dataset introduced in previous  study~\cite{pecorelli2019comparing}. The dataset contained over 8,500 manually validated code smells. Even with such high sounding number, the code smells formed a very small minority in the data. One example is the instance of God Class smell, which at its highest count in a release formed only around 1\% of all the classes present in that  release~\cite{pecorelli2019comparing}. To overcome this problem, the authors in~\cite{pecorelli2019role} tried five different data balancing methods, which were Class Balancer, resampling with balancing the dataset, SMOTE, Cost Sensitive Classifier and One Class Classifier. When compared to the BASELINE of no data balancing, the authors report highest performance overall with SMOTE, although there were differences between different smells. Our work is the first extensive empirical study to address the data imbalance for NLP-based SATD detection. We extend Pecorelli et al.,~\cite{pecorelli2019role} by evaluating multiple variants of SMOTE and include COST, Bagging, Boosting (classifier level ensemble techniques), and hybrid techniques for SATD detection from source code comments. We focus on the impact of balancing technique on evaluation metrics and provide a recommendation for each metric for both Within-Project and Cross-Project SATD detection from empirical results.

\section{Methodology}\label{sec:methodology}

In this section we describe the dataset used, discussed the various balancing techniques employed in our empirical study along with the experiment set up and evaluation criteria.

 \subsection{Machine learning classifiers}

Cruz et al.\cite{cruz2020detecting} empirically evaluated seven different machine learning algorithms for bad smell detection and found Random Forest and XGBoost achieved better overall performance. They have acknowledged the imbalanced data distribution in their data but have not employed explicit data balancing technique. This created undue advantage for the machine learning ensemble algorithms Random Forest and XGBoost which employ implicit balancing and are categorized under algorithm level (classifier level) balancing techniques, refer in \ref{classifier_balancing}. In our study, we include three machine learning classifiers:  Regression (Logistic Regression), Bagging (Random Forest) and Boosting (XGBoost). We choose Logistic regression for its superior performance in binary classification and its widespread usage in the past for NLP related tasks in software engineering \cite{genkin2007large,hall2011systematic,mantyla2018natural,cruz2020detecting}. Random Forest\cite{liaw2002classification}, is an ensemble algorithm (classifier level balancing technique explained in \ref{classifier_balancing}), based on the Bagging concept proposed by Breiman et al.\cite{breiman1996bagging}. It is aimed at reducing the variance associated with a model to increase the prediction capability. Random Forest has shown significant performance improvement in many software engineering tasks~\cite{cruz2020detecting,mustapha2019investigating,hossen2020hybrid,zampetti2017recommending, fontana2016comparing}. XGBoost, proposed by~\cite{chen2016xgboost} is a scalable, tree-based ensemble algorithm based on the Boosting concept proposed by Freung and Shapire~\cite{freung1997decision}.

\subsection{Balancing Techniques}

\begin{figure}[htbp]
\centerline{\includegraphics[width=0.5\textwidth]{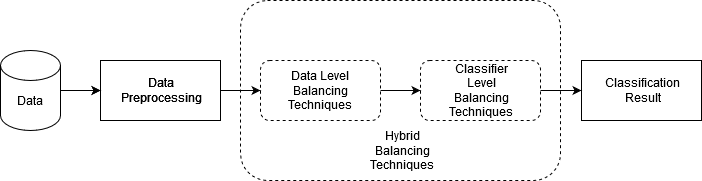}}
\caption{Balancing Techniques}
\label{fig:1}
\end{figure}

The balancing techniques could be broadly categorized into three groups which include Data Level techniques, Classifier Level techniques and Hybrid techniques, see Figure \ref{fig:1}.

\subsubsection{Data Level Techniques}

These techniques primarily focus on different data sampling schemes to reduce the skewness ratio in a dataset. Data Sampling is a host of techniques that transforms the distribution of the dataset with the objective of reducing the imbalance ratio in the dataset. Such schemes could be grouped into two categories, Oversampling and Undersampling. Oversampling is creating new samples from minority class either by mere duplication or by synthesizing samples. Undersampling is reducing the data samples from majority class by removing or selecting a subset of data samples. Multiple research including \cite{fitkov2013impact, kaur2018comparing, biswas2019exploring} establish the significance of oversampling over undersampling techniques. We include multiple variants of oversampling technique in our study.

\paragraph{SMOTE}
One of the widely used oversampling technique, Synthetic Minority Oversampling Technique (SMOTE) ~\cite{chawla2002smote}. In this, the minority class is over-sampled and new samples are formed based on randomly chosen samples that are close to each other. At first, random neighbours that are close to each other in the minority class are identified, then new samples are synthesized between any two neighbours and the same is replicated with other chosen neighbours. SMOTE has received a lot of success in data mining community. This success has been attributed to four factors, which are its simplicity, adaptability, generalizability and performance improvements~\cite{garcia2016tutorial}.

\paragraph{Borderline SMOTE}
Another variant of SMOTE, wherein data samples are not synthesized from the minority class randomly but with a specific target scheme. Borderline SMOTE (BLINE) was developed ~\cite{han2005borderline} with the focus to reduce the misclassification rate. The focus is on those data samples that are incorrectly classified and are lying near the ambiguous decision boundary of any two class X and Y. The target scheme is such that the data samples are synthesized near the decision boundary which segregates a class X from another class Y. 

\paragraph{ADASYN}
Adaptive Synthetic (ADASYN) sampling is a SMOTE-based over-sampling technique. The target scheme for this sampling technique is such that the new samples are synthesized in those areas where the density of the data samples is sparse, or where they are very few instances. The data samples in the high density regions are left undisturbed. This technique~\cite{he2008adasyn} has claimed to account for correct classification of difficult instances as it adaptively shifts the decision boundary.

\paragraph{SVMSMOTE}
SVMSMOTE(SVMSMT) is another SMOTE based over-sampling technique which employs SVM algorithm~\cite{boser1992training} to generate synthetic samples. The target scheme for synthesizing new samples of SVMSMOTE algorithm is focused around the hyperplane that separates each class.

\subsubsection{Classifier Level Techniques}\label{classifier_balancing}

Typically, the balancing schemes implemented inherently in the classification algorithms are referred to as classifier-level balancing techniques. Such techniques include, COST (assigning class weight) and a group of homogeneous classifiers (Bagging and Boosting) collectively working on the data samples for classification or prediction task. The ensemble algorithms Random Forest and XGBoost are explored as balancing techniques in \cite{ali2015classification,galar2011review,khoshgoftaar2015ensemble,haixiang2017learning}. Another classifier level technique is COST (or class weighted) in which higher weights are assigned to the class samples which are less in the training set so that it would penalize the model more during training for every incorrect prediction of the less represented class. This technique is simpler and efficient as it does not involve additional step of sample synthesis as in data sampling techniques. 

\subsubsection{Hybrid Techniques}
The hybrid balancing technique combines multiple balancing techniques to effectively handle the imbalance scenario. The ensemble technique which can handle the imbalance scenario is further augmented with either sampling or cost balancing techniques. In our study, we explore, Random Forest and XGBoost ensemble classifiers augmented with sampling and cost techniques. 

\subsection{Dataset}
For our experiments, we use the dataset created with 10 open source Java projects from Maldonado et al.~\cite{da2017using}. In their work, they extract and filter the source code comments from the 10 open source projects using heuristics including removal of license comments, task annotations, Javadoc comments and automatically-generated comments. The filtered comments were manually labelled and grouped into the following categories: Requirement Debt, Design Debt, Implementation Debt, Test Debt, Documentation Debt and Non-Technical Debt. We consolidated all the types of TD into a single TD class to facilitate binary classification. We have two classes: SATD and non-SATD class for simplicity and easier evaluation across several balancing techniques. Table~\ref{tab:datasetstatistics} lists the statistics of the source code comments data.


\begin{table}[htbp]
\begin{center}

\setlength\tabcolsep{2pt}
\caption{Dataset Statistics}
\begin{tabular}{|lcccccc|}
\hline
\bf \footnotesize{ Project} & \bf \footnotesize Release & \bf \footnotesize  \thead{\# of\\ Cmnts \cite{ren2019neural}}  & \bf \footnotesize  \thead{\# of\\ Cmnts\textsuperscript{*}}  & \bf \footnotesize SATD & \bf \footnotesize \thead{\% of\\ SATD }& \bf \footnotesize \thead{\% of\\ Non-SATD} \\ 
\hline
\bf \footnotesize Apache Ant &  \footnotesize 1.7.0 &  \footnotesize 4,137 & \footnotesize 2,992 &  \footnotesize 123 &  \footnotesize 4.11 &  \footnotesize  95.89 \\
\bf \footnotesize ArgoUML &  \footnotesize 0.34 &  \footnotesize 9,548 &  \footnotesize 4,725 &  \footnotesize 1,137 &  \footnotesize 24.06 &  \footnotesize  75.94 \\
\bf \footnotesize Columba &  \footnotesize 1.4 &  \footnotesize 6,478 & \footnotesize 3,842 &  \footnotesize 158 &  \footnotesize 4.11 &  \footnotesize  95.89 \\
\bf \footnotesize EMF &  \footnotesize 2.4.1 &  \footnotesize 4,401 & \footnotesize 2,382 & \footnotesize 81 &  \footnotesize 3.40 &  \footnotesize  96.60 \\
\bf \footnotesize Hibernate &  \footnotesize 3.3.2 &  \footnotesize 2,968 & \footnotesize 2,458 &  \footnotesize 423 &  \footnotesize 17.21 &  \footnotesize  82.79 \\
\bf \footnotesize JEdit & \footnotesize 4.2 &  \footnotesize 10,322 & \footnotesize 4,506 &  \footnotesize 228 &  \footnotesize 5.06 & \footnotesize  94.94 \\
\bf \footnotesize JFreeChart &  \footnotesize 1.0.19 &  \footnotesize 4,423 & \footnotesize 2,280 &  \footnotesize 106 &  \footnotesize 4.65 &  \footnotesize  95.35 \\
\bf \footnotesize JMeter &  \footnotesize 2.10 &  \footnotesize 8,162 & \footnotesize 4,029 &  \footnotesize 314 &  \footnotesize 7.79 &  \footnotesize  92.21 \\
\bf \footnotesize JRuby &  \footnotesize 1.4.0 &  \footnotesize 4,897 & \footnotesize 2,991 &  \footnotesize 458 &  \footnotesize 15.31 &  \footnotesize  84.69 \\
\bf \footnotesize Squirrel &  \footnotesize 3.0.3 &  \footnotesize 7,230 & \footnotesize 4,219 &  \footnotesize 216 &  \footnotesize 5.12 &  \footnotesize  94.88 \\
\hline

\bf \footnotesize Average &  \bf \footnotesize  &   \footnotesize 6,257 & \bf \footnotesize 3,442 &  \bf \footnotesize 324 & \bf  \footnotesize 9.08 &  \bf \footnotesize  90.91 \\
\hline
\end{tabular}
\label{tab:datasetstatistics}
\begin{tablenotes}
\item \footnotesize{\textsuperscript{*}After Duplicate Removal and Pre-processing in our study}
\end{tablenotes}
\vspace{-4mm}
\end{center}
\end{table}

\subsection{Processing Code Comments} \label{sec:processing}
We processed the code comments using following steps:
\begin{itemize}
    \item \textbf{Duplicate removal}: Duplicate comments are those that appear more than once in the training data. Example: '// Need to calculate this... just fudging here for now' source code comment appear three times in the training data. We identified such duplicates and removed 23,511 (37.75\%) duplicate source code comments from the training data used in Ren et al.'s~\cite{ren2019neural} neural network based SATD detection.
    \item \textbf{Special Character and Hyperlink Removal}: We removed special characters including \#, $@, \&,/, $',$", ( ),[],,{},$!, removed hyperlink references, truncated multiple spaces into single space and removed newline character.
    \item \textbf{Stop word removal}: We lower-cased the entire corpus, then stop-words including 'is', 'was', 'the', 'of', etc., which were very high in numbers and did not contribute to SATD feature detection were removed. We utilized the NLTK package~\cite{loper2002nltk} for removing the stop words from the dataset.
    \item \textbf{Lemmatization}: We employed WordNet Lemmatizer to convert the words to its respective base form depending on the context.
    \item \textbf{Restricting hollow comments}: We removed comments with only words such as 'ff', 'hhh', etc., that does not convey any meaning and included only those comments whose length is greater than 3 characters. 
\end{itemize}

The actual dataset employed by Ren et. al,~\cite{ren2019neural} for SATD classification had 62,566 source code comments. After performing all the pre-processing as discussed previously, the dataset used in our study consists of 34,424 source code comments, from which 3,248 were categorized as SATD. The percentage of SATD instances across projects ranges between 3.40\% to 24.06\%. On average each project has 3,442 source code comments, of which 324 (9.08\%) are SATD comments. 

\subsection{Experiment Configuration}
\paragraph{System Setup}
Logistic Regression, Random Forest and XGBoost models are implemented with Python based Scikit-Learn~\cite{pedregosa2011scikit} library. 
All the oversampling algorithm implementation is based on python based imbalanced-learn API \cite{JMLR:v18:16-365} with the sampling rate as 1.0 to ensure even class split between SATD and non-SATD classes. All experiments are executed as batch jobs in a HPC environment equipped with NVIDIA Tesla V100 GPU with 16 GB of memory.

\paragraph{Experiment Setup}
We replicated the experiment set up from Ren et al.~\cite{ren2019neural} to evaluate and compare with COST based Convolutional Neural Network. They used 90\% of stratified data samples from each open source project for training. Further, from the 90\% data, a stratified split of 10\% of data are used for cross-validation and 80\% is used for training the model. The data instances were stratified based on their label indicating SATD instances or not.  The trained model is evaluated on the remaining 10\% unseen test data in each open source project. The same experiment configuration is replicated for all the classifiers studied in this paper for Within-Project setup. For Cross-Project, for each of the 10 target projects, we used data from 9 projects as training data and evaluated the trained model on the target project which serves as the test data. We performed 10 fold cross-validation for both Within and Cross project setup.

\subsection{Evaluation Criteria}
We include four evaluation metrics for all our experiments, which include Precision, Recall, F1 and ROC-AUC scores. All the mentioned metrics are calculated based on the number of True Positive (TP), True Negative (TN), False Positve (FP) and False Negative (FN). TP indicate the number of correctly classified SATD comments, TN indicate the number of correctly classified non-SATD comments, FP indicate the number of incorrectly classified non-SATD comments, and FN indicate the number of incorrectly classified SATD comments.

\paragraph{Precision} 
In our experimental context, this metric represents the ratio of those comments that are precisely classifed as SATD comments  among all the comments subjected to prediction including both SATD and non-SATD comments. It is calculated as:
\begin{center}Precision $=\frac{TP}{TP+FP}$
\end{center}

\paragraph{Recall} 
This metric represents the ratio of actual SATD comments that are correctly classified as SATD comments. It is also known as the true positive rate or sensitivity of the classifier. It is calculated as:
\begin{center}Recall $= \frac{TP}{TP + FN}$\end{center}

\paragraph{F1} This metric represents the accuracy of the classifier based on the both precision and recall. It is calculated as:
\begin{center}F1 $= 2\times\frac{P \times R}{P + R}$
\end{center}
It is the harmonic mean of precision and recall which would account for the variance of each metric respectively. 

\paragraph{ROC-AUC} This metric represents the overall ability of the classifier to differentiate between SATD and non-SATD comments. This value represents the area under the Receiver Operating Characteristic curve which is based on the True Positive Rate and False Positive rate of the classifier. Typically, the value ranges between 0.5, being the lowest and 1.0, being the highest. The higher the ROC-AUC value, the better is the classifier ability to differentiate between SATD and non-SATD comments.

\section{Study  Results}\label{sec:results}
Here we tabulate and evaluate the impact of multiple balancing techniques for each metric, including Precision, Recall, ROC-AUC, and F1. We compare the balancing techniques against the BASELINE method with the same classifier without COST or Sampling balancing techniques. The data used for all these evaluations are based on clean data after duplicates removal as mentioned in section~\ref{sec:processing}.

\subsection{RQ 1 Precise detection of SATD comments  }
Here, our objective is to determine the balancing technique that consistently contribute to more precise detection of SATD comments in highly imbalanced data. Tables~\ref{tab:prec_results} lists the Precision scores for both Within and Cross-Project set up respectively.

\begin{table}[htbp!]
\caption{Consolidated Results: Within-Project and Cross-Project Precision scores}
\begin{center}
\begin{threeparttable}
\setlength\tabcolsep{6pt}
\begin{tabular} {l|lcccc}
\hline
\multicolumn{1}{c}{SCHEME} \\
\hline
\multicolumn{1}{c}{Within-Project}&
\multicolumn{1}{c}{}&
\multicolumn{1}{c}{}&
\multicolumn{2}{c}{HYBRID}&
\multicolumn{1}{c}{} \\
\hline
 &  \bf METHOD & \bf LR & \bf RF & \bf  XGB  & \bf Avg.\\
 \cline{2-6}
 &\bf BASELINE & \underline{0.710}  & 0.898 &  0.839\rlap{${^{w}}$}   & 0.816\\
\hline

\thead{\textbf COST\\SENSITIVE} & \bf WEIGHTED & 0.743\rlap{${^{b}}$}   &  0.904 & 0.788  & 0.812 \\
\hline
\multirow{4}{*}{ \thead{ \textbf DATA\\ SAMPLING}} & \bf SMOTE & 0.779  &  0.904 & 0.851\rlap{${^{w}}$}   & 0.845\\
& \bf BLINE & 0.825\rlap{${^{bw}}$}  & 0.902  &  0.816   & 0.848\\
& \bf ADASYN & 0.776 & \bf 0.905 & 0.804  & 0.828\\
& \bf SVMSMT & 0.850\rlap{${^{bw}}$}  &  0.895  & 0.835\rlap{${^{w}}$}  & 0.860 \\
\hline
& \bf Avg. & 0.780 & 0.901 &  0.822  & \\
\hline
\multicolumn{1}{c}{Cross-Project} \\
\hline

 &\bf BASELINE & 0.885\rlap{${^{ws}}$}  &  \bf 0.893\rlap{${^{s}}$} &  0.863\rlap{${^{ws}}$}   & 0.880\\
\hline
\thead{\textbf COST \\SENSITIVE} &  \textbf{WEIGHTED }&\underline{ 0.623}   & 0.881\rlap{${^{s}}$} & 0.713  & 0.739\\
\hline
\multirow{4}{*}{  \thead{\textbf DATA \\ SAMPLING}} & \bf SMOTE & 0.690 & 0.788 & {0.764}   & 0.747\\
& \bf BLINE & 0.724\rlap{${^{w}}$}  & 0.825 &  0.810   & 0.786\\
& \bf ADASYN & 0.683  & 0.744 &    {0.793}  & 0.740\\
& \bf SVMSMT &      {0.719\rlap{${^{w}}$}} & 0.854  & 0.811 & 0.795 \\
\hline
& \bf Avg. & 0.721 & 0.831 &  0.788  & \\
\hline
\end{tabular}
\label{tab:prec_results}
\begin{tablenotes}
\item \scriptsize{Highest score is bolded and lowest score is underlined.}
\item \scriptsize b,w,s - \scriptsize{Statistically significant change from baseline, weighted, and sampling respectively based on Wilcoxon-Signed Rank test with 95\% confidence level}
\end{tablenotes}
\end{threeparttable}
\end{center}
 \vspace{-6mm}
\end{table}

In Within-Project set up, the data sampling balancing technique SVMSMT/BLINE has improved the BASELINE and COST overall.
However, the BASELINE ensemble classifier Random Forest has significantly improved precision scores of COST and Sampling techniques of other classifiers. For cross-project SATD detection, the BASELINE has achieved the highest precision score while the COST and the data sampling techniques decreases the precision. 
Although the BASELINE scores appear almost equivalent, the ensemble classifier Random Forest has the lowest number of FP per TP with 0.11, followed by Logistic Regression with 0.13. Overall, neither COST nor Sampling consistently improve the precision scores over BASELINE across classifiers. The BASELINE ensemble classifier Random Forest contributes more to the precise detection of SATD comments in both Within and Cross-Project setup. However, between COST and sampling technique, the sampling techniques SVMSMT/BLINE has significantly improved the precision scores across all the classifiers over COST, in both Within and Cross-Project setup.

\setlength{\fboxrule}{2pt}
\begin{flushleft}
\fbox{\begin{minipage}{24em}
 \bf Recommendation: For consistent precise detection of SATD comments from imbalanced data, use classifier level Bagging ensemble technique (Random Forest). 
\end{minipage}}
\end{flushleft}
\vspace{6pt}

\subsection{RQ 2 Extensive detection of SATD comments}
Our motivation is to determine the balancing technique that consistently contribute to extensive detection of SATD comments in highly skewed data. Tables~\ref{tab:recall_results} lists the Recall scores for both Within and Cross-Project set up respectively. 

\begin{table}[htbp!]
\caption{Consolidated Results: Within-Project and Cross-Project Recall scores}
\begin{center}
\begin{threeparttable}
\setlength\tabcolsep{6pt}
\begin{tabular} {l|lcccc}
\hline
\multicolumn{1}{c}{SCHEME} \\
\hline
\multicolumn{1}{c}{Within-Project}&
\multicolumn{1}{c}{}&
\multicolumn{1}{c}{}&
\multicolumn{2}{c}{HYBRID}&
\multicolumn{1}{c}{} \\
\hline
 &\bf METHOD & \bf {\ \ LR} & \bf \ \ RF & \bf XGB  & \bf \ Avg.\\
 \cline{2-6}
&\bf BASELINE & \underline{0.277}  & 0.580  & 0.611  & 0.489\\
\hline
\thead{\textbf COST\\SENSITIVE} & \bf WEIGHTED & \bf 0.735\rlap{${^{b}}$}   & 0.586 & 0.730\rlap{${^{bs}}$}  & 0.684\\
\hline
\multirow{4}{*}{\thead{\textbf DATA\\ SAMPLING }} & \bf SMOTE & 0.726\rlap{${^{b}}$} & 0.653\rlap{${^{bw}}$}  & 0.688\rlap{${^{b}}$}  & 0.689\\
& \bf BLINE & 0.704  & 0.619 &  0.641   & 0.655\\
& \bf ADASYN & 0.731\rlap{${^{b}}$}  &  0.609\rlap{${^{b}}$} & 0.678\rlap{${^{b}}$}  & 0.673\\
& \bf SVMSMT & 0.658  & 0.644 & 0.644  & 0.649 \\
\hline
& \bf Avg. & 0.638  & 0.615 & 0.665 &  \\
\hline
\multicolumn{1}{c}{Cross-Project} \\
\hline
&\bf BASELINE & \underline{0.502}  & 0.617  & 0.657  & 0.592\\
\hline
\thead{ \textbf COST\\SENSITIVE} & \bf WEIGHTED & \bf0.754\rlap{${^{b}}$}   & 0.590 & 0.723\rlap{${^{b}}$}  & 0.689\\
\hline
\multirow{4}{*}{\thead{\textbf DATA\\ SAMPLING }} & \bf SMOTE & 0.727\rlap{${^{b}}$} & 0.674\rlap{${^{bw}}$}  & 0.702\rlap{${^{b}}$}  & 0.701\\
& \bf BLINE & 0.697  & 0.662 &  0.683   & 0.681\\
& \bf ADASYN &  0.730\rlap{${^{b}}$}  &  0.662\rlap{${^{bw}}$} & 0.697\rlap{${^{b}}$}  & 0.696\\
& \bf SVMSMT & 0.676  & 0.640 & 0.683  & 0.666 \\
\hline
& \bf Avg. & 0.681  & 0.641 & 0.691 &  \\
\hline
\end{tabular}
\label{tab:recall_results}
\begin{tablenotes}
\item \scriptsize{Highest score is bolded and lowest score is underlined.}
\item \scriptsize b,w - \scriptsize{Statistically significant change from baseline and weighted respectively based on Wilcoxon-Signed Rank test at 95\% confidence level}
\end{tablenotes}
\end{threeparttable}
\end{center}
\vspace{-6mm}
\end{table}

The Logistic Regression with COST balancing technique achieved the highest Recall scores in both Within and Cross-Project setup with 0.735 and 0.754 respectively. The sampling techniques SMOTE and ADASYN has consistently improved the Recall scores for all three classifiers over the BASELINE in Within and Cross-Project setup. Although the COST technique has recorded the highest recall scores in both the Within and Cross-Project setup, the improvement is not statistically significant over SMOTE and ADASYN sampling techniques. Further, SMOTE and ADASYN consistently improved the BASELINE with a statistically significant change while the COST technique is not consistent across all three classifiers.

\begin{flushleft}
\fbox{\begin{minipage}{24em}
\bf{ Recommendation: For consistent extensive detection of SATD comments from imbalanced data, use SMOTE/ADASYN sampling technique for both Within and Cross-Project setup. }
\end{minipage}}
\end{flushleft}
\vspace{6pt}

\subsection{RQ 3 Improving distinction capability of classifier for SATD detection}
Here our objective is to determine the balancing technique that contributes more to ROC-AUC score. ROC-AUC characterizes the distinctive capability of the machine learning model between technical debt and non-technical debt classes. Tables~\ref{tab:auc_results} lists the Recall scores for both Within and Cross-Project set up respectively. 

\begin{table}[htbp!]
\caption{Consolidated Results: Within-Project and Cross-Project AUC scores}
\begin{center}
\begin{threeparttable}
\setlength\tabcolsep{6pt}
\begin{tabular} {l|lcccc}
\hline

\multicolumn{1}{c}{SCHEME}&
\multicolumn{1}{c}{}&
\multicolumn{1}{c}{}&
\multicolumn{2}{c}{HYBRID}&
\multicolumn{1}{c}{} \\
\hline
\multicolumn{1}{c}{Within-Project} \\
\hline
 &\bf METHOD & \bf {LR} & \bf RF & \bf XGB  & \bf \ Avg.\\
 \cline{2-6}
 &\bf BASELINE & \underline{0.636} & 0.786 & 0.799   & 0.740\\
\hline
\thead{\textbf COST\\SENSITIVE} & \bf WEIGHTED & 0.855\rlap{${^{b}}$} &  0.790 &  \bf{0.857}\rlap{${^{b}}$}  & 0.834\\
\hline
\multirow{4}{*}{\thead{\textbf DATA\\ SAMPLING} } & \bf SMOTE & 0.854\rlap{${^{b}}$} & 0.822\rlap{${^{bw}}$}  & 0.838\rlap{${^{b}}$}  & 0.838\\
& \bf BLINE & 0.844  & 0.806  &  0.814  & 0.821\\
& \bf ADASYN & 0.856\rlap{${^{b}}$}  & 0.818\rlap{${^{b}}$}  & 0.832\rlap{${^{b}}$} &  0.835\\
& \bf SVMSMT & 0.822  & 0.786  & 0.816  & 0.808 \\
\hline
& \bf Avg. & 0.811  & 0.804 & 0.826   & \\
\hline
\multicolumn{1}{c}{Cross-Project} \\
\hline
&\bf BASELINE & \underline{0.747} & 0.804 & 0.824   & 0.792\\
\hline
\thead{\textbf COST\\SENSITIVE} & \bf WEIGHTED & \bf0.857\rlap{${^{b}}$}  &  0.790 &  0.850\rlap{${^{b}}$}  & 0.832\\
\hline
\multirow{4}{*}{\thead{\textbf DATA\\ SAMPLING }} & \bf SMOTE & 0.849\rlap{${^{b}}$} & 0.829\rlap{${^{bw}}$}  & 0.840  & 0.839\\
& \bf BLINE & 0.837  & 0.825  &  0.835  & 0.832\\
& \bf ADASYN & 0.849\rlap{${^{b}}$}  & 0.823\rlap{${^{w}}$}  & 0.840 &  0.837\\
& \bf SVMSMT & 0.827  & 0.814  & 0.835  & 0.825 \\
\hline
& \bf Avg. & 0.828  & 0.814 & 0.838   & \\
\hline
\end{tabular}
\label{tab:auc_results}
\begin{tablenotes}
\item \scriptsize{Highest score is bolded and lowest score is underlined.}
\item \scriptsize b,w - \scriptsize{Statistically significant from baseline, weighted respectively based on Wilcoxon-Signed Rank test with 95\% confidence level}
\end{tablenotes}
\end{threeparttable}
\end{center}
\vspace{-6mm}
\end{table}

The COST balancing technique with Logistic Regression/XGBoost recorded the highest ROC-AUC scores in both Within and Cross-Project setup.  
The sampling techniques SMOTE/ADASYN consistently improved the BASELINE ROC-AUC scores in both Within and Cross-Project SATD detection. The COST, on the other hand, is not consistent in improving the BASELINE across all the classifiers. Further the top score achieved with COST is not statistically significant over sampling techniques SMOTE/ADASYN. It is evident from Table~\ref{tab:recall_results} and Table~\ref{tab:auc_results} that, the COST decreases the recall and ROC-AUC scores for Random Forest, while the sampling techniques SMOTE/ADASYN improves them.

\begin{flushleft}
\fbox{\begin{minipage}{24em}
\bf Recommendation: For consistent better distinctive capability of the classifier (AUC-ROC) from imbalanced data, use SMOTE in both Within and Cross-Project setup. 
\end{minipage}}
\end{flushleft}
\vspace{6pt} 

\subsection{RQ 4 Balancing Precision and Recall for SATD Detection}

Here, our motivation is to determine the balancing technique that has consistently contributed more towards higher F1 score. Tables~\ref{tab:f1_results} lists the Recall scores for both Within and Cross-Project set up respectively. 

\begin{table}[htbp!]
\caption{Consolidated Results: Within-Project and Cross-Project F1 scores}
\begin{center}
\begin{threeparttable}
\setlength\tabcolsep{6pt}
\begin{tabular} {l|lcccc}
\hline
\multicolumn{1}{c}{SCHEME} \\
\hline
\multicolumn{1}{c}{Within-Project}&
\multicolumn{1}{c}{}&
\multicolumn{1}{c}{}&
\multicolumn{2}{c}{HYBRID}&
\multicolumn{1}{c}{} \\
\hline
 &\bf METHOD & \bf {LR} & \bf RF & \bf XGB  & \bf \ Avg.\\
 \cline{2-6}
 &\bf BASELINE & \underline{0.361} & 0.670 & 0.694   & 0.575\\
\hline
\thead{\textbf COST \\SENSITIVE} & \bf WEIGHTED & 0.730\rlap{${^{b}}$} &  0.680 &  \bf0.755\rlap{${^{b}}$}  & 0.722\\
\hline
\multirow{4}{*}{\thead{\textbf DATA\\ SAMPLING }} & \bf SMOTE & 0.748\rlap{${^{b}}$} & 0.738\rlap{${^{bw}}$}  &  0.753\rlap{${^{b}}$}  & 0.746\\
& \bf BLINE & 0.749  & 0.706  &  0.710  & 0.722\\
& \bf ADASYN & 0.750\rlap{${^{b}}$}  & 0.703  & 0.730 &  0.728\\
& \bf SVMSMT & 0.730  & 0.660  & 0.717  & 0.702 \\
\hline
& \bf Avg. & 0.678  & 0.704 & 0.726   & \\
\hline
\multicolumn{1}{c}{Cross-Project} \\
\hline
&\bf BASELINE & \underline{0.612} & 0.705\rlap{${^{w}}$} & \bf0.729   & 0.682\\
\hline
\thead{\textbf COST \\SENSITIVE} & \bf WEIGHTED & 0.674\rlap{${^{b}}$} &  0.679 &  0.712  & 0.688\\
\hline
\multirow{4}{*}{\thead{\textbf DATA \\ SAMPLING }} & \bf SMOTE & 0.700\rlap{${^{bw}}$} & 0.713\rlap{${^{w}}$}  & 0.723  & 0.712\\
& \bf BLINE & 0.700\rlap{${^{bw}}$}  & 0.718\rlap{${^{w}}$}  &  0.726  & 0.715\\
& \bf ADASYN & 0.697  & 0.708  & 0.721 &  0.709\\
& \bf SVMSMT & 0.684  & 0.711  & 0.728  & 0.708 \\
\hline
& \bf Avg. & 0.678  & 0.706 & 0.723   & \\
\hline
\end{tabular}
\label{tab:f1_results}
\begin{tablenotes}
\item \scriptsize{Highest score is bolded and lowest score is underlined.}
\item \scriptsize b,w - \scriptsize{Statistically significant from baseline, weighted respectively based on Wilcoxon-Signed Rank test with 95\% confidence level}
\end{tablenotes}
\end{threeparttable}
\end{center}
\vspace{-6mm}
\end{table}

XGBoost with COST and BASELINE XGBoost have recorded top F1 scores in Within and Cross-Project setup with 0.755 and 0.729 respectively. In Within-Project setup, the SMOTE balancing technique has consistently improved the F1 scores over BASELINE for all the classifiers, while COST is not consistent across classifiers. Further, the top F1 score with COST is not statistically significant over the F1 scores with BLINE/SMOTE. In Cross-Project setup, although the Boosting based ensemble classifier BASELINE XGBoost achieved the top Cross-Project F1 score of 0.729, it is not statistically significant over sampling techniques or over COST. The sampling techniques BLINE/SMOTE have a statistically significant improvement over BASELINE for Logistic Regression. For the ensemble classifiers Random Forest and XGBoost, the sampling technique have improved over COST and achieved an equivalent performance as that of the BASELINE.  Overall, the results show that the sampling techniques BLINE/SMOTE are more consistent in improving the F1 scores for all the classifiers in Cross-Project setup.
\begin{flushleft}
\fbox{\begin{minipage}{24em}
 \bf Recommendation: For consistent higher balance of Precision and Recall (F1-score), use SMOTE for Within-Project and BLINE/SMOTE for Cross-project SATD detection.

\end{minipage}}
\end{flushleft}
\vspace{6pt}

\subsection{Comparison to prior work}
Here we compare the F1 scores of our consistent top-performing model in both Within and Cross-project setup with previous works. 
First, we compare against the state of the art results from cost-sensitive CNN by Ren et al.~\cite{ren2019neural}. Then, we evaluate the performance improvement over Huang et al.~\cite{huang2018identifying} that employs feature selection as balancing technique for ensemble of Naive Bayes (NB) Multinomial classifiers.
For identical comparison with CNN that had duplicate source code comments in the training data, we retained the same data as in~\cite{ren2019neural} containing duplicate source code comments unlike~\cite{huang2018identifying} in which the duplicate source code comments were removed during the data preprocessing.  Table~\ref{tab:f1_comparison} lists the F1 comparison scores with both "Duplicates" and "No Duplicates" scenarios.

\begin{table}[htbp!]
\caption{F1 comparison }
\begin{center}
\begin{threeparttable}
\setlength\tabcolsep{1pt}
\begin{tabular}{lcccc|cccc}
\hline
& \multicolumn{4}{c}{WITHIN-PROJECT}&\multicolumn{4}{c}{CROSS-PROJECT}\\
\hline
& \multicolumn{2}{c}{\thead{Duplicates\\  \cite{ren2019neural}\textsuperscript{*}}}&\multicolumn{2}{c}{\thead{No\\ Duplicates}}
&\multicolumn{2}{c}{\thead{Duplicates\\
\cite{ren2019neural}\textsuperscript{*}}} &  \multicolumn{2}{c}{\thead{No \\Duplicates}}
\\
\cline{1-9}
Project & \thead{CNN \cite{ren2019neural}} & \thead{Our\textsuperscript{2}} & NB\cite{huang2018identifying} &  \thead{Our\textsuperscript{2}} & \thead{CNN \cite{ren2019neural}} & \thead{Our\textsuperscript{3}}  & NB\cite{huang2018identifying} & \thead{Our\textsuperscript{3}} \\
\hline
Ant & 0.445 &  \bf 0.690  & - & 0.476  &  \bf  0.660  & 0.627 & - & 0.554 \\
ArgoUML & \bf  0.932  & 0.907 & 0.705 & \bf  0.846 & \bf  0.878 & 0.861 & 0.828 & \bf  0.891\\
Columba &  0.741 & \bf  0.936 & 0.732  & \bf  0.815 & \bf  0.852 & 0.823 & 0.801 & \bf  0.865\\
EMF & 0.532 & \bf 0.778 & -  & 0.696 & \bf  0.679  & 0.483 & - & 0.522\\
Hibernate & \bf  0.887 & 0.851 & 0.752 & \bf  0.937  & \bf  0.826  & 0.825 & 0.788 & \bf  0.822 \\
JEdit & 0.622 & \bf  0.623 &  \bf 0.619  & 0.457  & \bf  0.599 & 0.423 & \bf  0.518 & 0.433\\
JFreeChart &  0.795 & \bf  0.919 & 0.581 & \bf  0.815  & \bf  0.739 & 0.574 & 0.687 & \bf  0.724 \\
JMeter & \bf  0.867 & 0.857 & 0.751  & \bf  0.882  & \bf  0.828 & 0.790 & 0.781 & \bf  0.847 \\
JRuby & 0.881 & \bf  0.885 & 0.782 & \bf  0.921  & 0.863 & \bf  0.908 & 0.841 & \bf  0.891\\
SQuirrel & 0.813 & \bf  0.838 & 0.628 & \bf  0.681   & \bf  0.739 & 0.692 & 0.651 & \bf  0.737\\
\hline
Avg. (10 proj) &  0.752 & \bf 0.828 & -  & 0.753 & \bf 0.766 & 0.701 & - & 0.726\\
Avg. (8 proj) &  - & - & 0.693  & \bf 0.794 & - & - & 0.736 & \bf 0.780 \\
\hline
\% Improv. &  - & \bf 10.10 & - & \bf 14.57 & \bf - & -9.27 & - & \bf 5.45\\
\hline
\end{tabular}
\begin{tablenotes}[para,flushleft]
\textsuperscript{*}\footnotesize{ Same data set up as in \cite{ren2019neural} (with duplicate source code comments)}\\
\textsuperscript{2}\footnotesize{ consistent top performer in Within-Project setup: XGBoost with SMOTE}\\
\textsuperscript{3}\footnotesize{ consistent top performer in Cross-Project setup: XGBoost with BLINE}
\end{tablenotes}
\label{tab:f1_comparison}
\end{threeparttable}
\end{center}
\vspace{-6mm}
\end{table}
Our classical machine learning classifier XGBoost with SMOTE achieved an overall average of 0.828 for ten projects exceeding the overall detection performance by 10.10\% against cost-sensitive CNN~\cite{ren2019neural}. XGBoost with SMOTE sampling improved the within-project F1 scores of all the projects except ArgoUML, Hibernate and JMeter. 
It appears that the deep learning model is more accurate for projects that have higher percentage of SATD instances. 
For the other projects, the XGBoost with SMOTE sampling has better F1 scores indicating 
improved performance of machine learning algorithm over deep learning in Within-Project setup. While for cross-project SATD detection the deep learning CNN model outperformed our consistent top performer in Cross-Project setup XGBoost with BLINE model by 9.27\% characterizing the improved feature generalization of source code comments across projects by deep learning model. This hints that for Within-Project SATD detection, classical machine learning with data balancing is better approach than deep learning. 

Our consistent top performer outperforms the prior work's ~\cite{huang2018identifying} ensemble of Naive Bayes (NB) classifiers with feature selection balancing technique in both Within and Cross-Project set up by 14.57\% and 5.45\% respectively. All the improvements in the F1 scores are statistically significant based on Wilcoxon-Signed rank test.

\subsection{Impact of Duplicates}
Here we highlight the impact of duplicate source code comments in the training data. From the Table~\ref{tab:f1_comparison}, it appears that the inclusion of duplicates can either improve or reduce our models' performance. We can see that the inclusion of duplicates increases the F1 score from 0.753 to 0.828 in the Within-Project setup. At the same time, the inclusion of duplicates decreases the Cross-Project average F1 to 0.726 to 0.701. It is reasonable to think that within a project duplicates can be useful as similar commenting style or copy-paste commenting code can result in exact duplicates. On the other hand for cross-project learning, it appears that learning from duplicates of some other projects reduces performance as such comments are unlikely to be present in the target project. 

More detailed project level comparison shows that
all the F1 scores under "No Duplicates" have dropped except for the projects Hibernate, JMeter, and JRuby that have a higher number of duplicate SATD comments, with an average of 10\% of duplicate SATD comments.  The other remaining projects have less than 6\% of duplicate SATD comments. This somewhat surprising finding suggests that fewer duplicates below a particular threshold increase the detection performance. 
While on the other hand, removal of higher duplicates increases the prediction performance as observed in Hibernate, JMeter and JRuby projects. This indicate higher duplicates have a negative impact on the prediction score. 
Perhaps the feature learning over the same SATD features of duplicate source code comments in the training data does not enable the machine learning classifier to detect new SATD features when subjected to new source code comments in the test data.


In Cross-Project setup, each target project is trained with the data of the 9 other projects, and thus has a higher average of 778 duplicate SATD source code comments. This in contrast with the Within-Project setup's per project average of 87 duplicate SATD source code comments, is much higher. The removal of SATD duplicate comments has improved the overall F1 performance from 0.701 to 0.726 asserting our previous observation that higher duplicates have a negative impact on the prediction performance.

\section{SATD Feature Interpretability and Analysis}\label{sec:quality}

In this section, we 
interpret features from the trained machine learning classifiers and study the impact of the balancing techniques on SATD feature selection. 
There have been significant recent advances in explaining black-box ML models~\cite{strumbelj2010efficient, ribeiro2016should, NIPS2017_7062}. We employ the explainability framework SHAP (SHapley Additive exPlanations)~\cite{NIPS2017_7062}, due to its feature interpretation capability.
SHAP is based on cooperative game theory, where the goal is to predict which groups (coalitions) will emerge from pool of player to collect payoffs. In other words, we use SHAP to see features as players that form coalitions to detect SATD. 
SHAP uses Shapley value~\cite{shapley1953value}, and the Shapley value of a feature represents the average contribution of a feature with all possible coalitions.
With Shapley values, we identified the number of features that contributed to the prediction, features that have a high impact (features that have higher mean of absolute Shapley values) on the prediction, and the features that do not contribute to the prediction.

We choose the Logistic Regression classifier for JMeter project in Within-Project setup for a brief insight on the impact of balancing techniques for SATD detection. We chose JMeter as an example project as it has 7.79\% of SATD instances, close to the overall SATD average of 9.08\%. We can see in Table ~\ref{tab:within_proj_perf_analysis} that in the JMeter example the performance is in line with previous results. BASELINE has the highest precision but suffers from poor recall while balancing techniques sacrifice little precision to improve recall and therefore offer better overall performance in F1 and ROC-AUC. 

\begin{table}[htbp]
\begin{center}
\setlength\tabcolsep{6pt}
\caption{Logistic Regression's SATD Detection Performance for JMeter (Within-Project)  
}
\begin{tabular}{|ccccc|}
\hline
\bf \footnotesize{ Technique } &  \bf  \footnotesize Precision  &  \bf  \footnotesize  Recall &  \bf \footnotesize  F1 &   \bf  \footnotesize ROC-AUC   \\ 
\hline
\bf BASELINE  & \bf 0.875 &  0.389 &  0.538 & 0.692 \\
\bf COST & 0.784 &  \bf 0.806 &  0.795 & 0.892 \\
\bf SMOTE	  & 0.848 &  0.778 &  0.812 & 0.882 \\
\bf ADASYN   & 0.829 &  \bf 0.806 &  \bf 0.817 & \bf 0.895 \\
\bf BLINE  & 0.844 &  0.750 &  0.794 & 0.868 \\
\bf SVMSMT & 0.871 &  0.750 &  0.806 & 0.870 \\
\hline
\end{tabular}
\label{tab:within_proj_perf_analysis}
\end{center}
\vspace{-6mm}
\end{table}
Next, we look if balancing techniques differ in terms of features. 
Table~\ref{tab:shap_feature_stats} shows the number of features that are contributing or not contributing in each balancing technique, e.g., BASELINE has 875 contributing and 2116 non-contributing features. The table shows also how many new features each balancing technique introduces in comparison to BASELINE. COST echoes its inherent nature (of adjusting instance weight) and introduces 1.4\% of new features, while the sampling technique introduces new features between 12-15\%, reflecting its new sample synthesis ability. 



\begin{table}[htbp]
\begin{center}
\setlength\tabcolsep{1.4pt}
\caption{SHAP Feature Contribution Statistics (Within-Project JMeter)}
\begin{tabular}{|c|c|c|c|c|c|c|}
\hline
  \bf \footnotesize Technique &    \bf \footnotesize \thead{BASE-\\LINE} &     \bf \footnotesize COST  &   \bf \footnotesize  SMOTE &   \bf \footnotesize  ADASYN &     \bf \footnotesize SVMSMT &   \bf \footnotesize BLINE \\ 
\hline
\thead{\bf{ Contributing} \\ SHAP value $\neq 0$ } & 875 & 	870 & 	916 & 	905 & 	914 & 	919 \\
\thead{\bf{ Non-Contributing} \\ SHAP value $= 0$ }& 2116 & 	2121 & 	2075 & 	2086 & 	2077 & 	2072 \\

\bf New Feature count & - & 	12 & 	124 & 	108 & 	112 & 	140 \\
\bf  New Feature \% & - & 	1.4\% & 	13.5\% & 	11.9\% & 	12.2\% & 	15.2\% \\
\hline
\end{tabular}
\label{tab:shap_feature_stats}
\vspace{-2mm}
\end{center}
\end{table}

Feature count alone cannot tell us if the balancing techniques really are meaningfully different from BASELINE so we also look in to the top contributing features of each. Table~\ref{tab:top_features_per_technique} contains the top 10 features of SATD detection by each of the balancing technique.

\begin{table}[htbp]
\begin{center}
\setlength\tabcolsep{1.2pt}
\caption{Top 10 contributing features (JMeter Within-Project)}
\begin{tabular}{|cccccc|}
\hline
 \bf \footnotesize{ \thead{BASE-\\LINE} } &  \bf  \footnotesize COST  &  \bf  \footnotesize  SMOTE &  \bf \bf \bf \footnotesize  ADASYN &   \bf  \footnotesize SVMSMT & \bf  \footnotesize BLINE \\ 
\hline
todo & todo & todo & todo & todo & todo \\
hack & hack & hack & hack & perhaps & perhaps \\
file & later & appear & bug & 	appear & hack \\
nonjavadoc & file & used & used & doe & used \\
fix & 	helper & doe & 	need & cleaning & could \\
later & fix & 	nonnls & number & fix & 	exception \\
helper & one & 	exception  & appear & used & improve \\
yet & 	yet & 	improve & improve & disconnected & wrapped \\
encoding & nonjavadoc & perhaps & allow & nonnls & always \\
found & currently & later & reason & note & nonjavadoc \\
\hline
\end{tabular}
\label{tab:top_features_per_technique}
\end{center}
\vspace{-2mm}
\end{table}
The feature \textbf{'todo'} is the most contributing SATD feature across all the techniques. \textbf{'hack'} is the second most contributing feature across sampling techniques but does not list in the top 10 of SVMSMT. Similarity to BASELINE is the highest in COST with 8 shared features. Data sampling techniques have lower number of shared features with the BASELINE. SMOTE has 3 (todo, hack, later), ADASYN has 2 (todo, hack), SVMSMT (todo, fix), BLINE (todo, hack, nonjavadoc). So COST is the most similar to BASELINE also in terms of the top 10 contributing features while data sampling techniques have larger differences to BASELINE. When investigating feature importance, in a single project one would suspect that project specific words would have more weight. Indeed this is a difficulty others have experienced when using regression coefficients as measures of feature importance~\cite{rantala2020prevalence}. It appears that SHAP avoids this problem as only one project specific word \textbf{'nonnls'} is seen in the table. 

Finally, we explore the newly introduced features in comparison to the BASELINE to understand whether the new features make sense in SATD detection. Table~\ref{tab:top_introduced_features} contains the top 3 features which are new contributing features, i.e., shifted from non-contributing to contributing features by the application of balancing techniques. 
\begin{table}[htbp]
\begin{center}
\setlength\tabcolsep{5.2pt}
\caption{Top 3 introduced features (JMeter Within-Project)}
\begin{tabular}{|ccccc|}
\hline
  \bf  \footnotesize COST  &  \bf  \footnotesize  SMOTE &  \bf \bf \bf \footnotesize  ADASYN &   \bf  \footnotesize SVMSMT & \bf  \footnotesize BLINE \\ 
\hline
 validation & decision & although & decision & stopped \\
 initialization & important & generally & sensible & nothing \\
 fetched & anything & notused & 	extra & extra \\
\hline
\end{tabular}
\label{tab:top_introduced_features}
\end{center}
\vspace{-2mm}
\end{table}
Feature \textbf{'decision'} appears in two techniques and it could very well be used when discussing decisions made while coding. Also feature \textbf{'extra'} is in two techniques and it could be used referring to code having extra logic. COST technique appears to have words that are the most technical (\textbf{'validation'}, \textbf{'initialization'}, \textbf{'fetched'}), while data sampling techniques introduce new features that appear more appropriate in SATD detection like \textbf{'important'}, \textbf{'anything'}, \textbf{'nothing'}, \textbf{'sensible'}, \textbf{'notused'}, \textbf{'extra'}, \textbf{'decision'}.

\section{Discussion}\label{sec:discussion}
In this section, we discuss the factors affecting the choice of a balancing technique and the implications to practitioners and researchers. Our results show that no single balancing technique can provide consistent higher performance across multiple metrics.
Even though the sampling technique SMOTE has consistently improved Recall, ROC-AUC, and F1 scores of BASELINE across multiple classifiers, the absolute average difference between COST and SMOTE for the same metrics is less than 4\%. This suggests performance tradeoff as the crucial parameter for choosing between COST and SMOTE in which the SMOTE incurs an additional data synthesis step while the former does not.

Ensemble classifiers Random Forest and XGBoost handle class imbalance with their algorithmic design. Random Forest creates a random subset of samples to generate multiple decision trees while XGBoost creates more decision trees based on the misclassified samples in the training data. XGBoost's approach results in higher F1, ROC-AUC, and Recall while Random Forest results in higher Precision. However, the maintenance of Random Forest is more expensive with model training time being 2.5 times than that of the XGBoost.


The real-world application of AI/NLP techniques for SATD detection through source code comments depends heavily on time and resource availability. Limited time and resources could not allow a higher number of incorrectly classified SATD instances, warranting the precise detection of SATD comments. On the contrary, if enough time and resources are available to validate SATD comment classification manually, then extensive detection (recall) of SATD comments should be the ideal choice of metric. Our study would enable practitioners and researchers to choose an appropriate balancing technique for NLP-based SATD detection in Within-Project (limited data) and Cross-Project (ample data) setup. 

Another important implication for researchers is the removal of duplicate source code comments in the training data. Even though the duplicate source code comments accompanied with duplicate source code are to be treated as separate technical debt candidates, such duplicate comments should be avoided while training machine learning models as they affect the prediction scores.

\section{Threats to Validity}\label{sec:threats} 
The superior performance of classical machine learning algorithm XGBoost in Within-Project set up over Deep Learning algorithm Convolution Neural Network might be due to the limited training data size with an average of 3,442 source code comments. Even though the source code comments within a project will always be lesser than consolidated source code comments from multiple projects, the Within-Project setup's training data size might play a vital role in the classical machine learning algorithm's improved performance. The data sampling rate of all the evaluated sampling techniques have been evaluated at 1.0 (maximum), in our study, for realizing an even class distribution between technical debt and non-technical debt classes. The SATD detection performance of each evaluation metric is highly susceptible to change if the sampling rate changes.

The balancing technique recommendations from this study are based on the source code comments data with an average class imbalance ratio of 9.08\% between SATD and non-SATD classes. All the balancing techniques employed in this study have been evaluated for the class imbalance ratio of 1:10 approximately (for every one technical debt source code comment, there are ten non-technical debt source code comments). A higher or lower class imbalance ratio before applying a balancing technique might have a different impact on the evaluation metrics. More research should identify the threshold for the ideal class imbalance ratio for improved SATD detection performance. 

Another crucial factor that might impact the generalizability of our results is the choice of machine learning models. The applicability of our recommendations might vary for different machine learning classifiers.

\section{Conclusion}\label{sec:conclusion} In this study, we  investigated the effectiveness of multiple balancing techniques to detect SATD comments from imbalanced data. 
Our study is based on a manually labeled dataset from 10 open source projects.
We make four contributions.

First, we studied the performance of balancing techniques. We found that SMOTE provides the most consistent improvement for Recall, ROC-AUC and F1 in Within and Cross-Project SATD detection for all the classifiers included in this study. However, if one does not want the extra data synthesis step in SMOTE sampling, the classifier level COST (class weight) balancing technique is a potential alternate.
The ensemble (classifier level balancing technique) algorithm (Random Forest and XGBoost) achieved the best Cross-Project Precision and F1 scores in imbalanced data without explicit balancing technique such as COST or SMOTE. This shows that, for Cross-Project SATD detection in imbalanced data focusing on high Precision or high F1, the Bagging or Boosting ensemble techniques serves as a potential choice apart from sampling technique. 

Second, we compare our results to previous works
in SATD detection that include feature selection based Naive Bayes Multinomial~\cite{huang2018identifying} and COST based CNN~\cite{ren2019neural}. Our top-performing XGBoost with SMOTE/BLINE sampling technique outperforms Huang et al.'s~\cite{huang2018identifying} approach by 14.57\% and 5.45\% in both Within and Cross-Project setup respectively. Then, in comparison with the state of the art CNN, we find that our consistent top performer in terms of F1 score (XGBoost with SMOTE) beats CNN in Within-Project SATD detection, offering a 10\% improvement. However, in Cross-Project SATD detection, our top performer loses to CNN by 9\%. For SATD detection, it suggests that in a Within-Project setup, classical ML is a better choice. However, the deep learning approach shows improved generalization capability in cross-project setup. 

Third, our feature analysis shows that all techniques use a similar set of features with at least 85\% of features being the same in all techniques. Sampling techniques have a larger difference to BASELINE than to COST technique. Highly shared top features were the words "todo" and "hack". Sampling techniques also appear to be able to introduce new feature words that are sensible in the SATD context which further support their consistent improved performance over COST. 

Fourth, we developed a SATD classification tool based on this study that works in two modes, online and batch. Our web tool's batch mode would help reduce the initial efforts for labeling a huge volume of source code comments data.


In the future, we intend to apply and evaluate deep learning architectures for SATD detection from source code comments. In particular we are keen to explore the capability of BERT (a transfer learning scheme, that is already pre-trained on a huge corpus) for SATD detection in imbalanced data.
\section* {Acknowledgement}\label{sec:ack}
The authors acknowledge the financial support by the Academy of Finland (grant ID 328058) and computational infrastructure by CSC Finland.

\bibliographystyle{IEEEtran}
\bibliography{mybibfile}

\end{document}